\documentclass[aps,prl,preprint,showpacs,groupedaddress]{revtex4}
\usepackage{graphicx,epsfig,pstcol,pstricks}

\begin{document}
\title{
Non-Markovian melting: a novel procedure to generate initial liquid like phases for small molecules for use in computer simulation studies.
}
\author{Carl McBride}
\author{Carlos Vega}
\author{Eduardo Sanz}
\affiliation{
Departamento  de Qu\'{\i}mica F\'{\i}sica.
Facultad de Ciencias Qu\'{\i}micas. Universidad Complutense de Madrid.
Ciudad Universitaria 28040 Madrid, Spain.
}
\date{4 February 2005}
\begin{abstract}
Computer simulations of liquid phases require an initial configuration from which to begin.
The preparation of such an initial configuration or `snapshot' 
often involves the melting of a solid phase.
This melting is usually undertaken by heating the system at low pressure, followed
by a lengthy re-compression and cooling once the melt has formed.
This note looks at a novel technique to produce a 
liquid phase from a perfect crystal using a standard Monte Carlo simulation code. 
\end{abstract}

\keywords{Monte Carlo, Random Number Generators, Initial liquid configuration}

\pacs{02.70.Uu, 05.10.Ln, 61.20.Ja}

\maketitle

\section{Introduction}
An important prerequisite for the simulation of liquid phases
is the generation of a suitable initial configuration or `snapshot'.
Given that the density of such materials is often close to that
of the crystalline solid (or even higher in the case of water) this is not a trivial task.
Simply placing atoms or molecules into the simulation `box' in a 
haphazard fashion is almost always destined to fail; there being a
high probability of either overlap for `hard-core' systems,
or the generation of very high energy configurations for `soft' potentials.
The approach generally adopted to this situation is 
to start from a perfect crystalline structure 
(see \cite{AllenTildesleyChp5})
and then either heat the solid to beyond its melting point 
or simply expand the system to a low density state (this can be done
either by reduction of pressure or more simply by enlarging the 
simulation box) so that the solid melts. 
Once the system has melted (this melting process 
being judged perhaps by some sort of order parameter), the 
system is then compressed and/or cooled to the 
desired thermodynamic conditions and an often 
substantial equilibration run performed.
To complicate the situation further, it should be stated that 
in computer simulations the solid does not melt at the thermodynamic
melting temperature (at a given constant pressure) or at the equilibrium
melting pressure (at a given constant temperature). 
In simulation studies of bulk solid phases where no
free surface is present superheating (super-expansion) of the solid phase
is the rule rather than the 
exception \cite{JCP_2004_120_11640,MP_2005_103_0001,JCP_2002_116_08876,MS_2004_30_0131_nolotengo}. 
In fact the temperature at which a solid phase melts at constant
pressure is usually $20\%-30\%$ higher than the equilibrium melting 
point.  
For this reason 
free energy calculations are used to determine the equilibrium melting point
of a model $T_{m}$, where the pressure and chemical potential of both 
phases are identical. An alternative method is to create a solid-liquid interface 
and allow this to reach equilibrium \cite{JCP_2002_116_09352}.
Once a coexistence point is known, the rest of the melting curve can be traced out
using the Gibbs-Duhem integration technique \cite{JCP_1993_98_04149,CPC_2001_141_0403}.

Algorithms do exist to  produce an initial disordered system, 
such as the 
`Skew Start' method implemented in the Molecular Dynamics simulation code `{\it Moldy}' 
\cite{CPC_2000_126_0310}.
However, in this work a simple technique is presented that provides a 
rapid path to the production of liquid phases.
This technique can be applied to many Monte Carlo 
\cite{JASA_1949_44_0335,JCP_1953_21_01087}
simulation codes and requires
no changes to be made to the source code.

In general no computer simulation is truly ergodic, {\it i.e.} it does not have time
to visit {\it all} of the points in phase space.
However, one hopes that 
the duration of the simulation is such that the trajectory followed is representative of the system.
After a sufficiently long run one hopes that the system is in its equilibrium state and
ensemble averages yield correct (to within statistical uncertainty) values
for thermodynamic quantities.
Note that for complex systems a `sufficiently long' run may be very long indeed. 
Facing the problem of `broken ergodicity' has led to the development of 
special Monte Carlo (MC) methods, such as the Jump-Walking technique 
developed by Frantz, Freeman and Doll \cite{JCP_1990_93_02769}

In any MC simulation the quality of the random number generator 
(RNG) used  is of fundamental importance.
Producing a series of pseudo-random numbers from arithmetical methods
is a far from trivial task, and much effort has been devoted 
to this subject (see \cite{KnuthBookVol2}).
Indeed, given the exact solution for the $2-d$ Ising model,
Monte Carlo simulations have been used as a test of the quality of  random number generators
\cite{PRL_1992_69_003382,IJMPC_1994_5_0547_preprint},
a test which many so called `good' RNG's have failed.

In this paper we describe an interesting observation which, to the best of our
knowledge, has not been previously reported.
It has been found that when performing short, consecutive  Monte Carlo simulations, 
using the final configuration of the previous run as the initial
configuration of the new run, and maintaining the same initial seed 
throughout the consecutive Monte Carlo runs  
then solid phases melt even for temperatures below the melting point, 
$T_{m}$.  
The decay of the solid structure is due to the non-Markovian character of the Monte Carlo simulations when performed 
as described.
Although the Metropolis importance sampling scheme is used to accept the trial configurations,
the principle of detailed balance or
microscopic reversibility is not satisfied. 
A number of short consecutive  Monte Carlo runs is 
equivalent to periodically restarting the RNG from the same initial point during a simulation.
This resetting of the RNG breaks the Markov chain.

Here use is made of this ``non-Markovian" melting for practical purposes.
By taking a disrupted configuration, and by subjecting it to 
a standard equilibration run, 
results are produced that
agree very well with systems obtained via much more circuitous routes
involving many more simulation cycles.

Three examples are presented, a simple Lennard-Jones system, an ionic salt (NaCl), and the melting of ice-I$_h$
to liquid water.

\subsection{Simple system: The Lennard-Jones fluid}
A system of 256 atoms, 
interacting via the Lennard-Jones 12-6 (LJ) potential 
\cite{PPSL_1931_43_461_nolotengo}, 
were arranged in a face 
centered cubic close packed structure (see Fig. 1).
For the Lennard-Jones system the thermodynamic state is described in terms of reduced units \cite{AllenTildesleyApenB1}  
such that 
$\rho^{*}= (N/V) \sigma^{3}$, $p^*=p/(\epsilon/\sigma^3)$ and $T^{*}= T /(\epsilon/k)$ where 
$\epsilon$ and $\sigma$ are the parameters of the LJ potential, $N$
is the number of molecules (or atoms in this case) of the system, and $V$ is
the total volume.
For Canonical ensemble ($NVT$) simulations one  MC cycle includes one trial move per 
particle 
(either a translational move, or for non-spherical molecules, a rotational move).
For $NpT$ simulations  a trial change in the volume of the system is also performed. 
The pair potential was truncated at $r=2.7\sigma$, and standard long
range corrections to the energy were added.

It is often useful to quantify the degree of order in a 
system with a suitable order parameter. 
In this study the intensity of the Bragg reflection 
from the $hkl$ planes of the crystal structure is used: 
\begin{equation}
I_{hkl} = |F_{hkl}|^2 = F_{hkl} F_{hkl}^*
\label{sokI}
\end{equation}
which is given by the square of the structure factor
defined as:
\begin{equation}
F_{hkl} = \frac{1}{N} \sum_{i=1}^{i=N} f_i \exp \left( 2 \pi i ( hx_i + ky_i + lz_i)\right)
\label{sok}
\end{equation}
where $x_i$, $y_i$ and $z_i$ are coordinates of molecule $i$ relative to 
the vectors that define the simulation box.
The atomic scattering factor, $f_{i}$
was arbitrarily set to one.
For $hkl$ the planes with the most intense line were chosen.
For the perfect face centered cubic solid  $I_{hkl}=1$, and for a isotropic liquid
$I_{hkl}=0$.

The LJ solid was studied by performing $NpT$ simulations
at $T^{*}=0.75$ and $p^*=1.00$.
Under these conditions the thermodynamically stable phase is the solid 
\cite{MP_1995_85_0023_nolotengo,MP_1995_85_0043_nolotengo}.
As an illustration of this stability an $NpT$ simulation was performed.
After 200000 cycles the average value of $I_{hkl}$ of the Monte Carlo
run was $I_{hkl}=0.72$ and the density was $\rho^{*} =0.98$.
A snapshot of this final configuration
is shown in Fig. 2.  
In contrast to this situation, 20 consecutive runs of 10 $NVT$ Monte Carlo cycles
are performed at the same temperature, also starting from the perfect crystal structure with $\rho^{*}=0.98$.
Each final configuration of a Monte Carlo run becomes the initial configuration for
the subsequent run, whilst maintaining the same initial seed for the RNG.
The result of this brief process of only 200 MC cycles is presented
in Fig. 3.
The evolution of $I_{hkl}$ as
a function of the number of cycles is presented in Fig. 4 for both
the non-Markovian melting and a standard $NpT$ run.  
The structure factor decays rapidly, having $I_{hkl} = 0.001$ after 200 MC cycles. 
The result is dramatically different from that of the 
plateau reached by the standard run; the initial crystal structure is now completely disrupted.
This disrupted configuration was then equilibrated for $15 \times 10^3$ cycles in a standard $NpT$ MC run. 
The density and internal energy 
obtained for the supercooled liquid  is 
$\rho^* = 0.8779$ and $U/NkT=-8.3216$,
which compares extremely well with 
$\rho^* = 0.8782$ and $U/NkT=-8.3236$
for the liquid phase obtained by melting the solid at high temperatures 
and then slowly cooling the system back down  to $T^{*}=0.75$ and $p^*=1.00$.
\subsection{Ionic system: Simulation of NaCl}
In this section an ionic system is studied in a similar fashion to 
that of the Lennard-Jones described in the previous section.
The system comprised of 512 ions, half sodium and half chlorine, with the Fm$\overline 3$m space group (Fig. 5).
The parameters for this model are taken from Ref.
\cite{JCP_1994_100_03757}. 
The ions consist of a  
LJ potential plus a Coulombic charge, either $e$ or $-e$, located at the 
center of the ion. 
The melting point for this model was calculated to be $T_{m}=1304$ K at $p=1$ bar \cite{SanzVegaUnpublished}
by means of free energy  
calculations using the Frenkel-Ladd method for the solid phase \cite{JCP_1984_81_03188}.
The LJ potential is truncated at $r= 10.6 $~\AA~ and long range
corrections to the energy were accounted for. Electrostatics were
treated using the Ewald sum technique \cite{AdPL_1921_64_0253_nolotengo}.
With this in mind non-Markovian melting was undertaken 
at a temperature of
$T=1200$ K and $p=1$ bar.
In principle, at this temperature and pressure the solid is the stable phase, 
and this was indeed the case during a standard $NpT$ MC run of 200000 cycles
(see Fig. 6). The resulting structure factor is $I_{hkl} = 0.748$ and $\rho=1.7$ g/cm$^3$
(note that $I_{hkl}^{\rm perfect} = 1$).
However, after 10 consecutive $NVT$ Monte Carlo runs of 10 cycles 
($T=1200$ K , $\rho=1.7$ g/cm$^3$ ), 
again repeating the RNG seed as in the LJ case, $I_{hkl}$ drops to 0.02.
The corresponding snapshot of this structure is shown in Fig. 7.

Once again, the disordered configuration obtained from
non-Markovian melting was equilibrated in the $NpT$ ensemble for $5 \times 10^3$ cycles,
followed by a production run of $15 \times 10^3$ cycles.
This resulted in a system with a density of $\rho = 1.28$ g/cm$^3$
and an internal energy of $U=-171.3$ Kcal/mol.
This compares very well with  $\rho = 1.28$ g/cm$^3$ and $U=-171.4 $ Kcal/mol 
obtained for a supercooled system of NaCl obtained from the standard
route ({\it i.e.} heating the solid until it melts and then cooling it slowly).

In Fig. 8 the results of changing the number of MC cycles in NVT
runs before repeating the seed 
is presented. These results were obtained for  
$T=1200$ K , $\rho=1.7$ g/cm$^3$.
It is interesting to note that non-Markovian melting occurs for runs of up to 60 cycles.
However, for runs of 75 cycles the effect is reduced to a slight drop in $I_{hkl}$.
\subsection{Molecular system: Simulation of supercooled water}
In this example we present the case of a molecular fluid; water.
To describe water the TIP4P \cite{JCP_1983_79_00926} model was used. This model  consists of 
a LJ site located on the oxygen atom, two positive charges located on 
the hydrogen atoms, and a negative charge is located $0.15$~\AA~ from the oxygen along the bisector of the H-O-H
angle. 
The LJ potential was truncated at $r= 8.5$~\AA~ and long range
corrections to the energy were accounted for. 
Electrostatics were
treated using the Ewald sum technique.
The TIP4P is one of the most popular models of water used in 
biological simulations. The melting temperature at $p=1$ bar of ice I$_h$ 
for this model has been determined recently to be 232 K 
\cite{JCP_2000_112_08534,JCP_2004_121_07926,PRL_2004_92_255701}.
The `normal' path to producing a system of super-cooled water would be to 
take a crystalline water structure, typically ice I$_h$ \cite{bookPhysIce}, and melt it at
a high temperature. For the TIP4P model a `high' temperature would be one 
in excess of 310 K 
\cite{MP_2005_103_0001,JCP_2002_116_08876}.
Once the system had melted it would then be cooled to 230 K
involving a substantial period for equilibration.
In this example the system simulated consists of 432 TIP4P 
\cite{JCP_1983_79_00926,JCP_2004_121_01165,PRL_2004_92_255701}
water molecules in the ice-I$_h$ crystal structure.

As before, two simulations were performed.
In the first simulation a standard MC run of 200000 cycles was undertaken in the $NpT$ ensemble
at a temperature of $T=230$ K and $p=1$ bar, yielding a density of $\rho=0.94$ g/cm$^3$. At this temperature and pressure the solid is 
the thermodynamically stable phase.
From this final configuration, 20 consecutive runs of 10 $NVT$ Monte Carlo cycles each 
({\it i.e.} 200 cycles in total) are performed, and as before,
each simulation was initiated from the output configuration of the previous run, whilst maintaining
the RNG seed the same in each case.

In Fig. 9 we see the result of a single  run of 200000 Monte Carlo cycles 
and in Fig. 10 we see the result of 20 runs of 10 Monte Carlo cycles.
In the case of the 200000 cycles simulation we see that the crystal 
lattice has remained largely unchanged apart from small displacements about the 
mean positions of the molecules with  $(I_{hkl} = 0.284)$. 
This is what one should expect since we
are simulating below the melting temperature of the model.
However, in contrast we can see that after only 20 runs of 10 Monte Carlo cycles
the crystal structure is all but lost. 
This structure-less system was then simulated for $1.2 \times 10^5$  conventional $NpT$ MC cycles. 
This resulted in a system with a density of $\rho = 0.99$ g/cm$^3$
and an internal energy of -11.01 Kcal/mol.
This compares very well with 
a density of $\rho = 1.00$ g/cm$^3$
and an internal energy of -10.98 Kcal/mol obtained via a standard heating-melting-cooling  simulation route.

\section{Conclusion}
By using a short period RNG ({\it i.e.} a number of consecutive, very short 
simulations maintaining the same initial RNG seed) 
it is possible to rapidly disrupt the crystal structure, even below the melting
temperature of the model under consideration. This phenomena is a 
result of the non-Markovian character of the simulations.
Once this disordered configuration has been obtained (typically 
within 100-200 cycles, less than 1 minute of CPU on a standard 
personal computer) 
it is the possible
to perform a standard simulation Monte Carlo to obtain an equilibrated
supercooled liquid. 
In the examples in this work a `short' period is between $\approx 15000$
and $\approx 30000$ random numbers {\it i.e.} 10 cycles of 6 random numbers (particle choice,
choice of move, 3 displacements which can be either of
translational or of rotational type 
and acceptance) for 
250 - 500 molecules.

The methodology has been tested for three different systems, the simple 
Lennard-Jones system, the ionic NaCl model and the TIP4P model of water.
Advantages of this method for liquids is that it is not necessary to 
raise the temperature of the system, thus avoiding the creation 
of high energy molecular conformations.
Another feature is that it is independent of the RNG and it is not necessary to modify the source of
the simulation code, which may not always be available.

It is worth noting that once a disordered state  has been formed it is very rare
to observe re-crystallization in simulation studies, this requiring 
the activated process of nucleation. As an example of this the 
re-crystallization of water to ice I$_h$ has only ever been seen once 
during a computer simulation \cite{N_2002_416_00409}.
\section{Acknowledgments}
\begin{acknowledgments}
This research has been funded by project
FIS2004-06227-C02-02 
of
the Spanish DGI (Direccion General de Investigacion).
One of the authors, C. M., would like to thank the Comunidad de Madrid
for the award of a post-doctoral
research grant (part funded by the European Social Fund). E. S. would like to thank the Spanish Ministerio de
Educacion for the award of an FPU grant.
\end{acknowledgments}
\bibliography{bibliography.bib,local.bib}
\clearpage
\begin{list}{}{\leftmargin 2cm \labelwidth 1.5cm \labelsep 0.5cm}

\item[\bf Fig. 1] Caption of Figure 1.
Snapshot of the perfect lattice of Lennard-Jones atoms. ($I_{hkl} = 1.00$)

\item[\bf Fig. 2] Caption of Figure 2.
Snapshot of Lennard-Jones after 200000 $NpT$ MC cycles. 
Results for $T^{*}=0.75$, $p^{*}=1.00$. The average value of 
the density and of the order parameter obtained from this NpT 
run are  $\rho^{*}=0.98$ and $I_{hkl} = 0.72$ respectively.

\item[\bf Fig. 3] Caption of Figure 3.
Snapshot of the disordered Lennard-Jones after only 20 runs of 10 $NVT$ Monte Carlo cycles.  
The runs were performed at  $T^{*}=0.75$,  $\rho^{*}=0.98$.
The order parameter of the snapshot is 
 $I_{hkl} = 0.001$.

\item[\bf Fig. 4] Caption of Figure 4.
Plot of the decay of $I_{hkl}$ for the LJ system with respect to the number of Monte Carlo cycles. 
Dashed line for 20 runs of 10 $NVT$ MC cycles ($T^{*}=0.75$,  $\rho^{*}=0.98$) compared with 
 the first 500 MC cycles of a standard $NpT$ run (solid line) performed 
at $T^{*}=0.75$,  $p^{*}=1.00$ (after 200000 cycles $I_{hkl} = 0.72$)

\item[\bf Fig. 5] Caption of Figure 5.
Snapshot of the perfect lattice of NaCl. ($I_{hkl} = 1.00$)

\item[\bf Fig. 6] Caption of Figure 6.
Snapshot of NaCl after 200000 $NpT$ MC cycles at 1200 K and 1 bar. The average density
and translational order parameter obtained from the run is  $\rho=1.7$ g/cm$^3$
and $I_{hkl} = 0.748$ respectively.

\item[\bf Fig. 7] Caption of Figure 7.
Snapshot of the disordered NaCl after only 10 runs of 10 $NVT$ Monte Carlo steps at 1200 K and 
$\rho=1.7$ g/cm$^3$. The value of the translational order parameter of the snapshot is
$I_{hkl} = 0.02$.

\item[\bf Fig. 8] Caption of Figure 8.
Plot of $I_{hkl}$ with respect to the number of MC cycles for contiguous block of 10, 50, 60, 65 and 75 cycles along with a  standard $NVT$ MC simulation for NaCl at $T=1200K$ and $\rho=1.7$ g/cm$^3$.

\item[\bf Fig. 9] Caption of Figure 9.
Snapshot of ice-I$_h$ after 200000 $NpT$ Monte Carlo steps at 230 K and 1 bar. The hexagonal lattice is still evident. The average value of the density and of the translational order 
parameter obtained in the run is $\rho=0.94$ g/cm$^3$ and $I_{hkl} = 0.284$ respectively.

\item[\bf Fig. 10] Caption of Figure 10.
Snapshot of the ice-I$_h$ after 20 runs of 10 $NVT$ Monte Carlo steps at 230 K and 
$\rho=0.94$ g/cm$^3$ . The translational order parameter of the final snapshot is 
$I_{hkl} = 0.002$.

\end{list}

\clearpage
 \begin{figure}[!]
 \includegraphics[height=400pt,width=500pt]{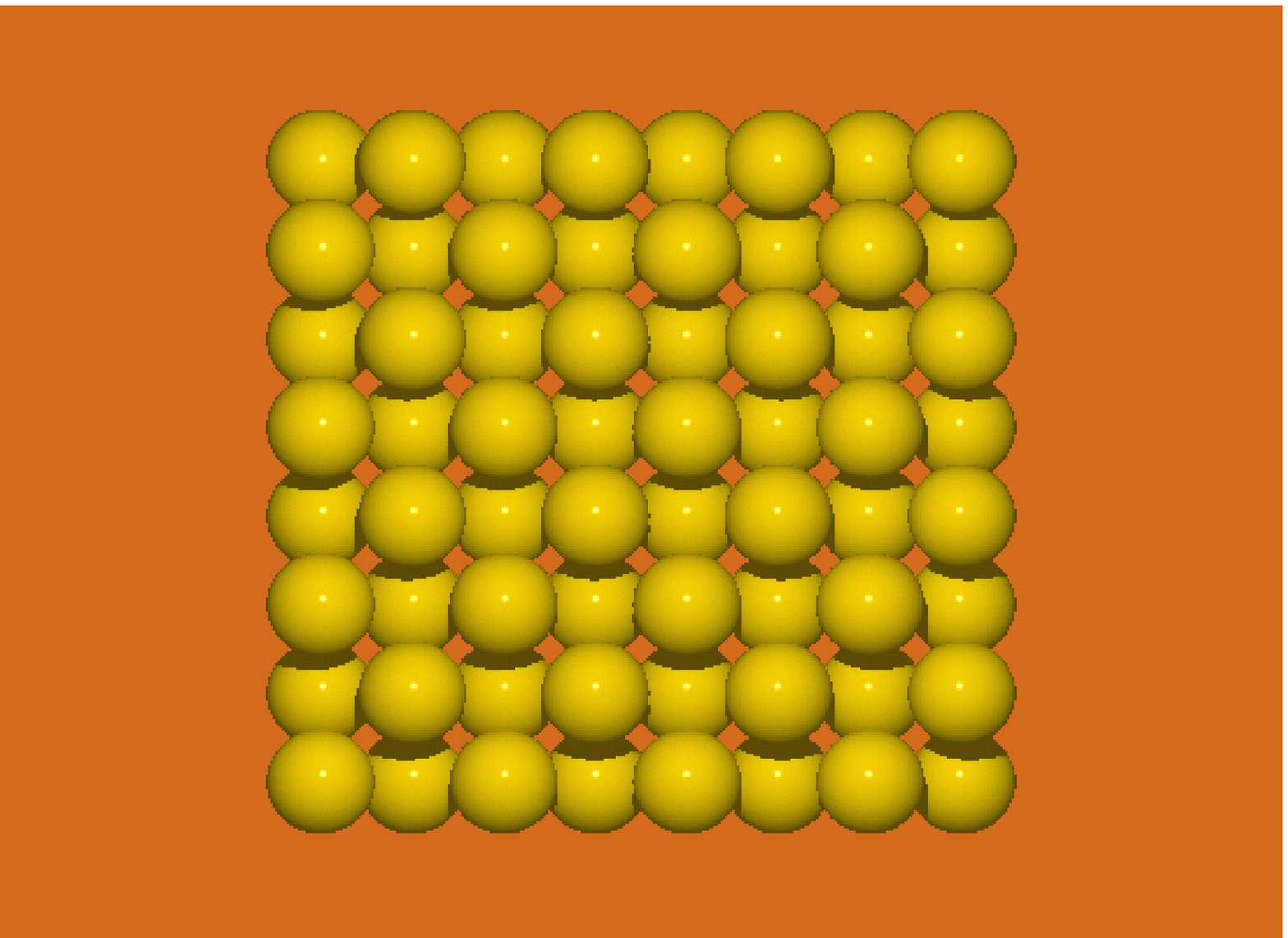}%
 \caption{
}
 \end{figure}
 \begin{figure}[!]
 \includegraphics[height=400pt,width=500pt]{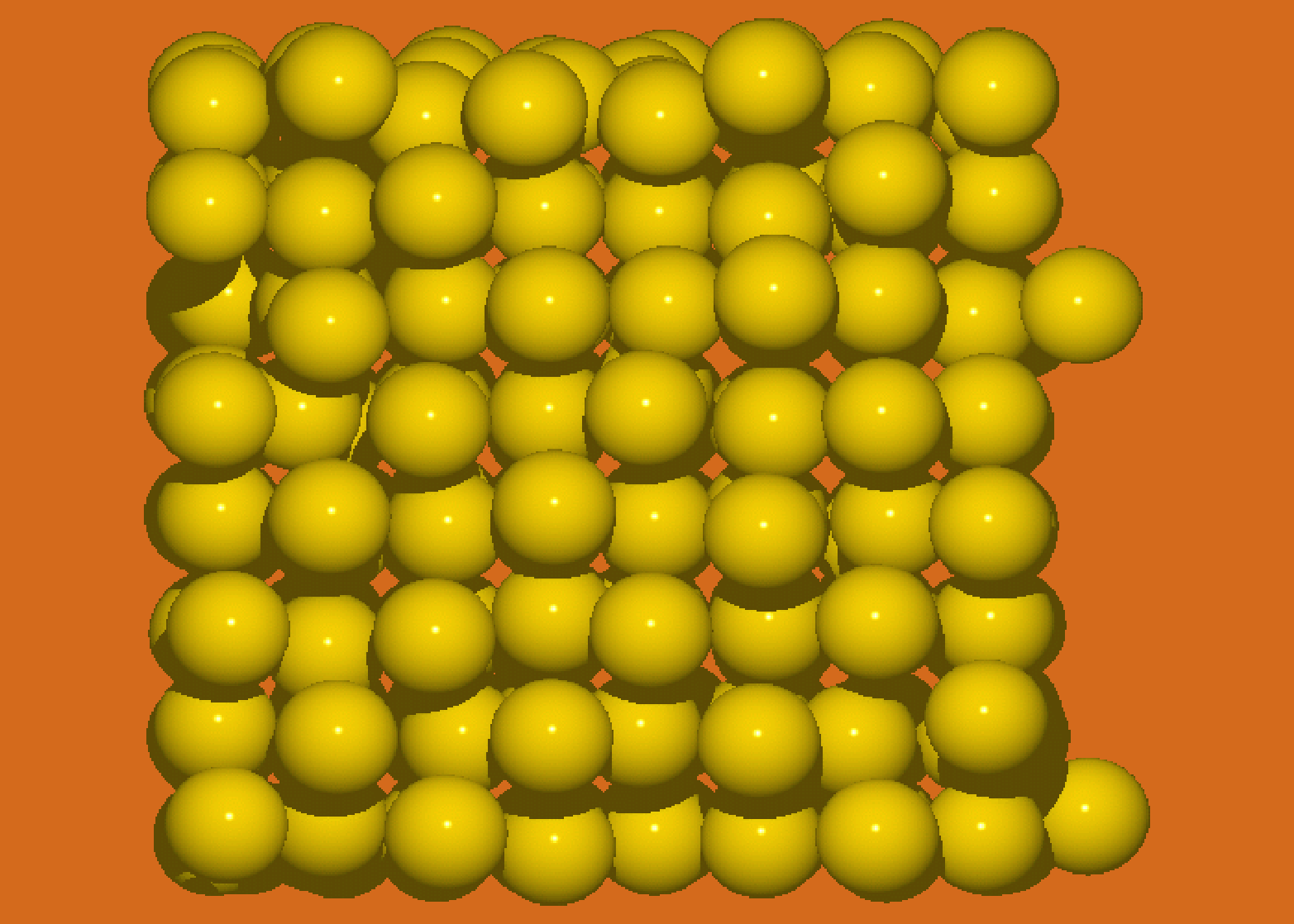}%
 \caption{
}
 \end{figure}
 \begin{figure}[!]
 \includegraphics[height=400pt,width=500pt]{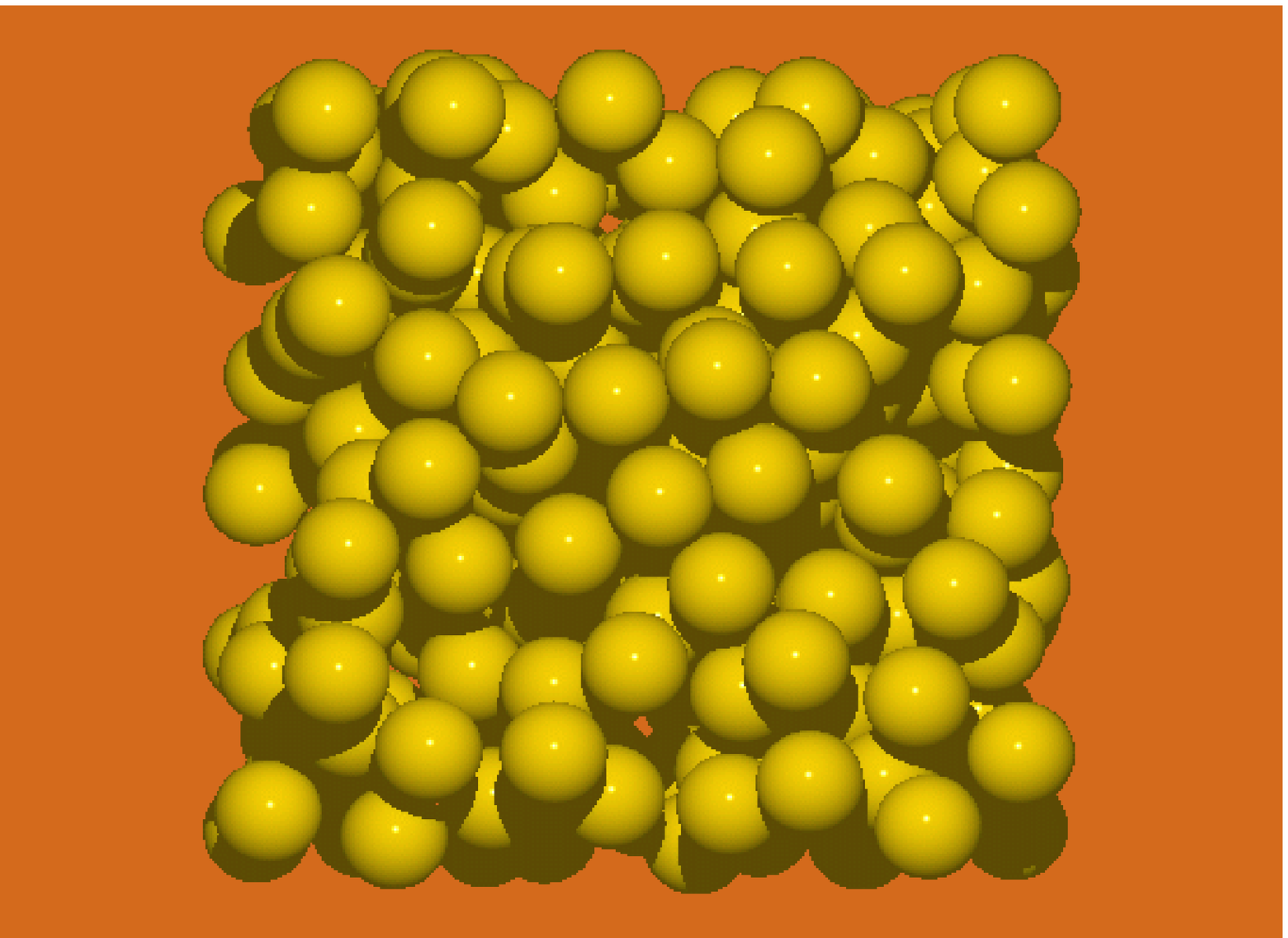}%
 \caption{
}
 \end{figure}
 \begin{figure}[!]
 \includegraphics[height=500pt,width=400pt,angle=-90]{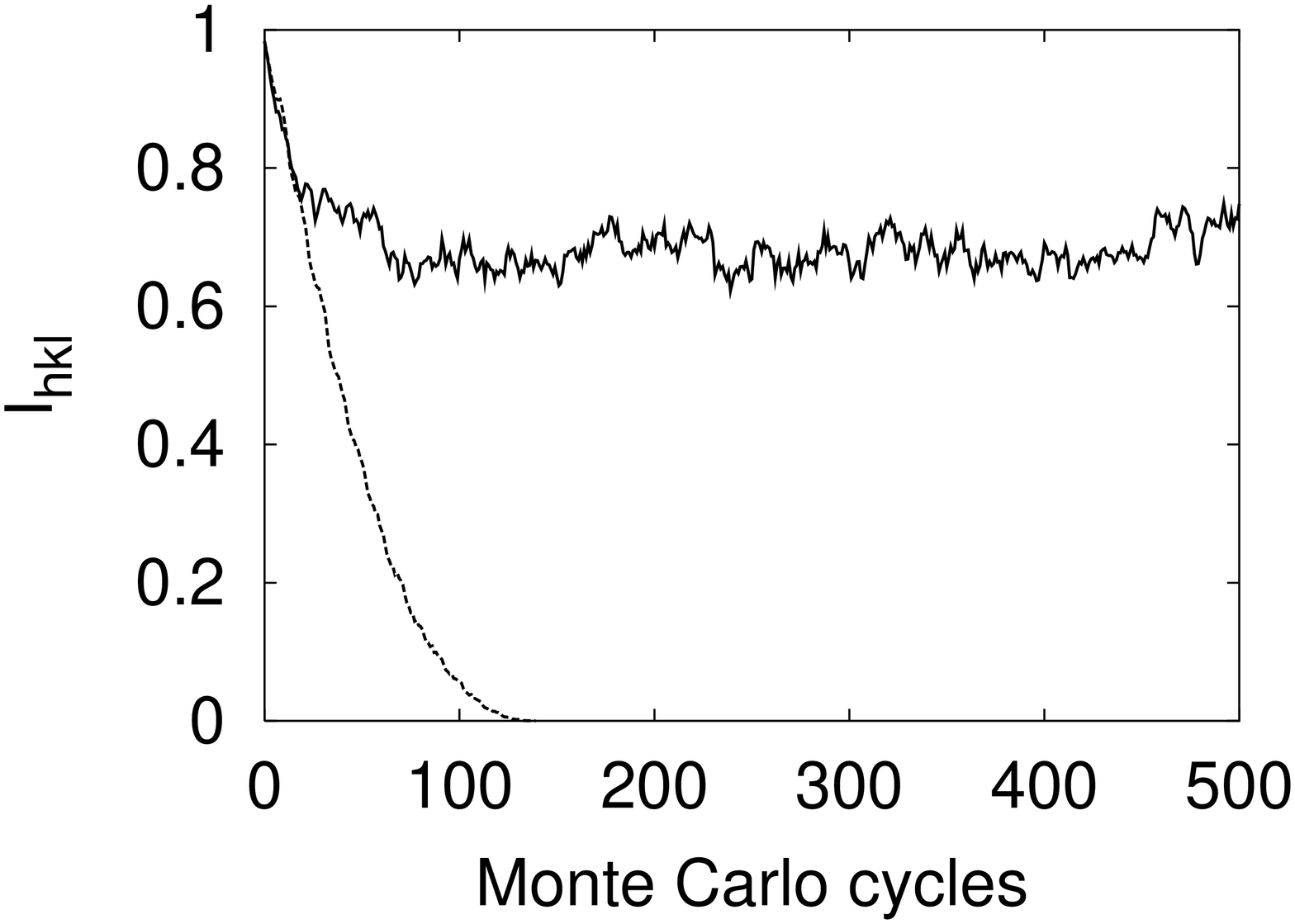}%
 \caption{
}
 \end{figure}
 \begin{figure}[!]
 \includegraphics[height=400pt,width=500pt]{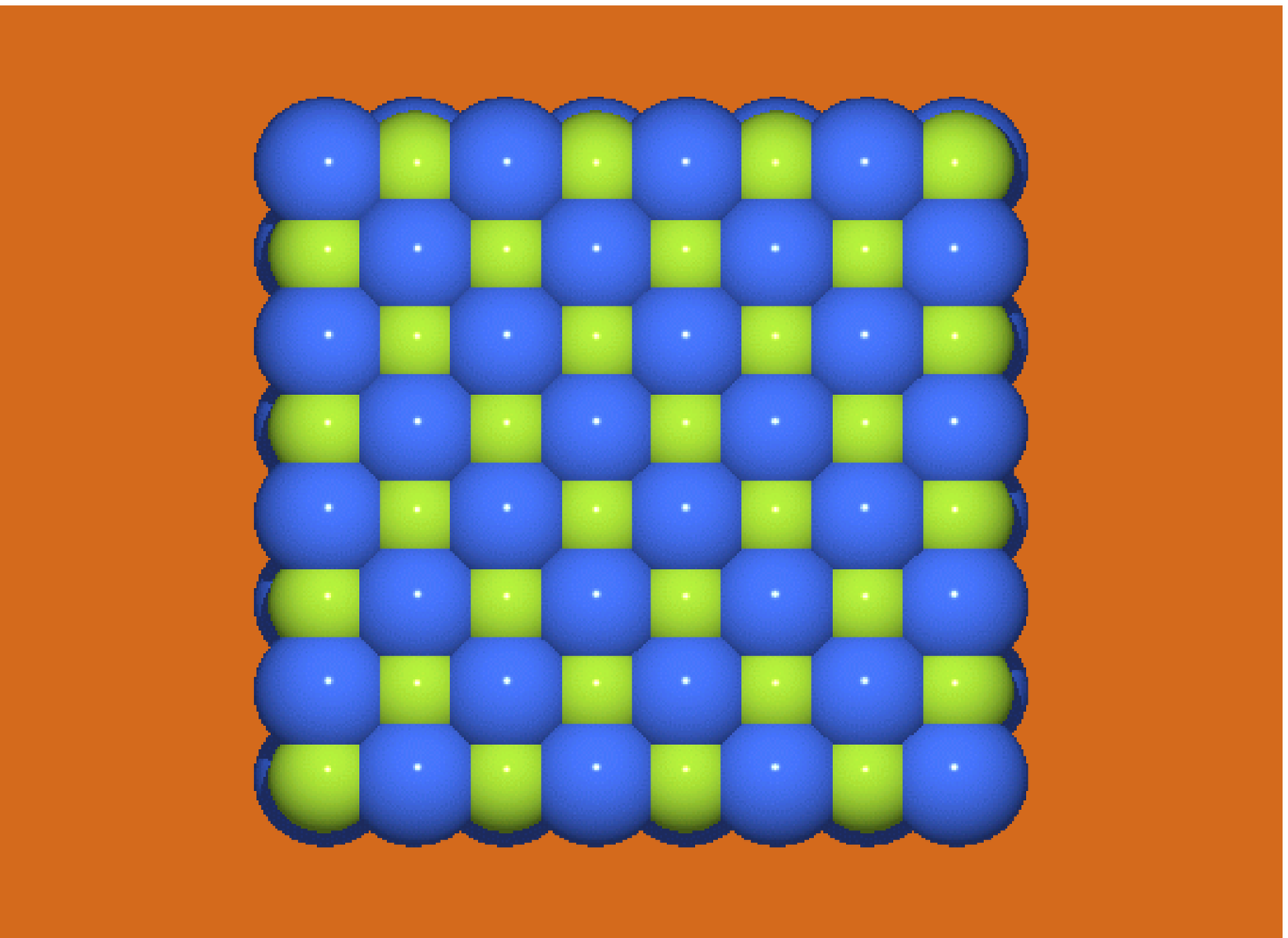}%
 \caption{
}
 \end{figure}
 \begin{figure}[!]
 \includegraphics[height=400pt,width=500pt]{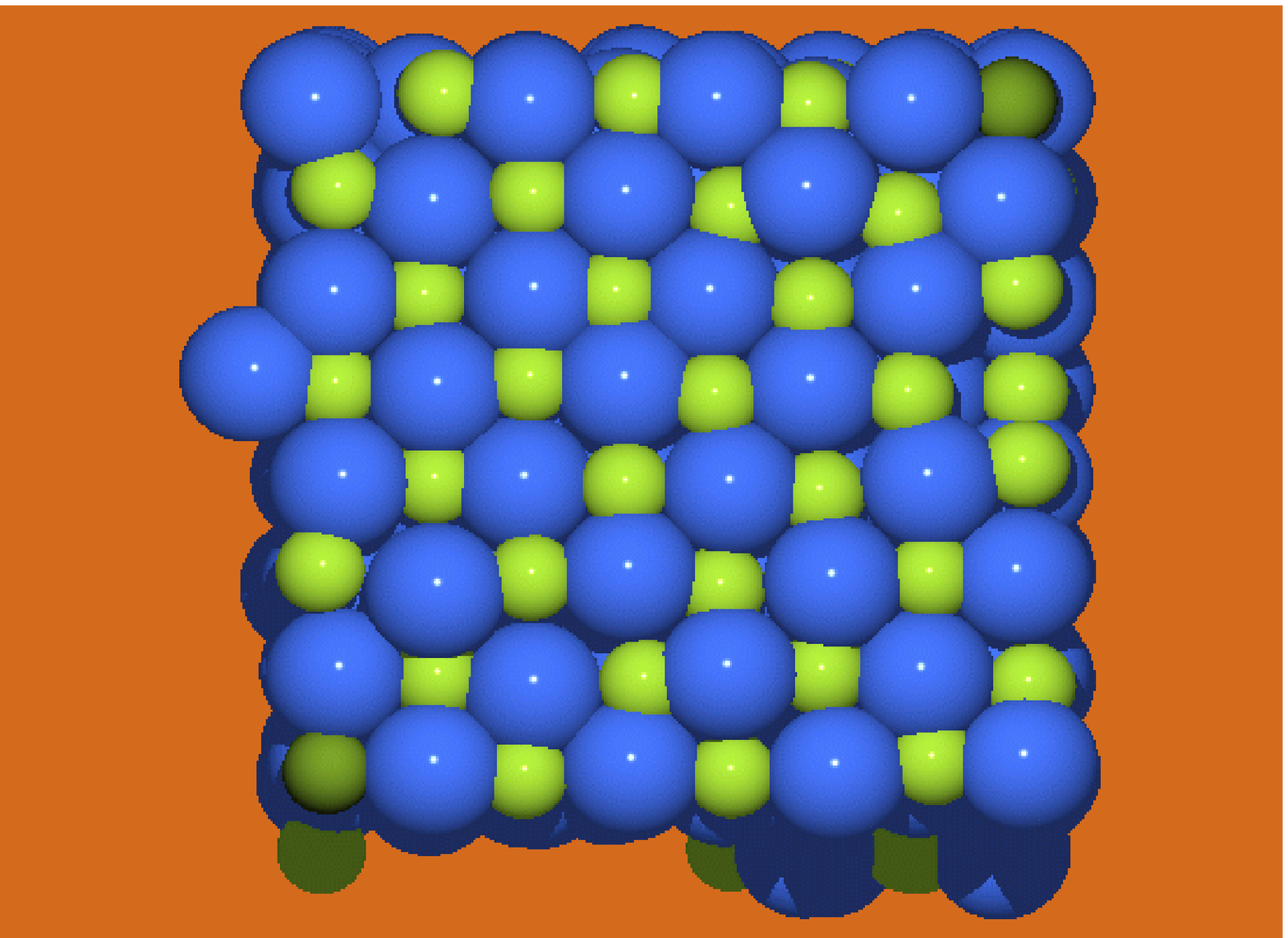}%
 \caption{
}
 \end{figure}
 \begin{figure}[!]
 \includegraphics[height=400pt,width=500pt]{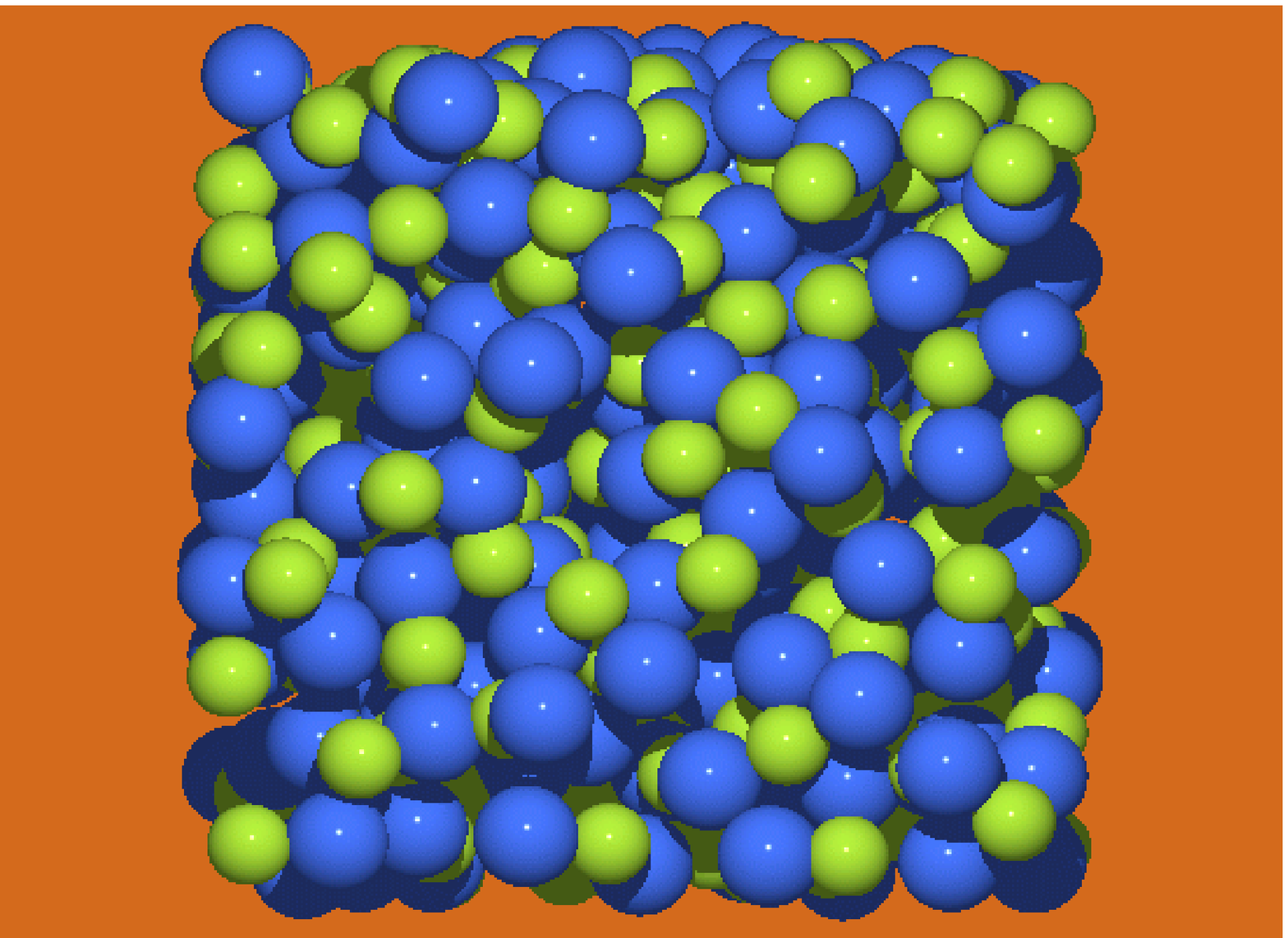}%
 \caption{
}
 \end{figure}
 \begin{figure}[!]
 \includegraphics[height=500pt,width=400pt,angle=-90]{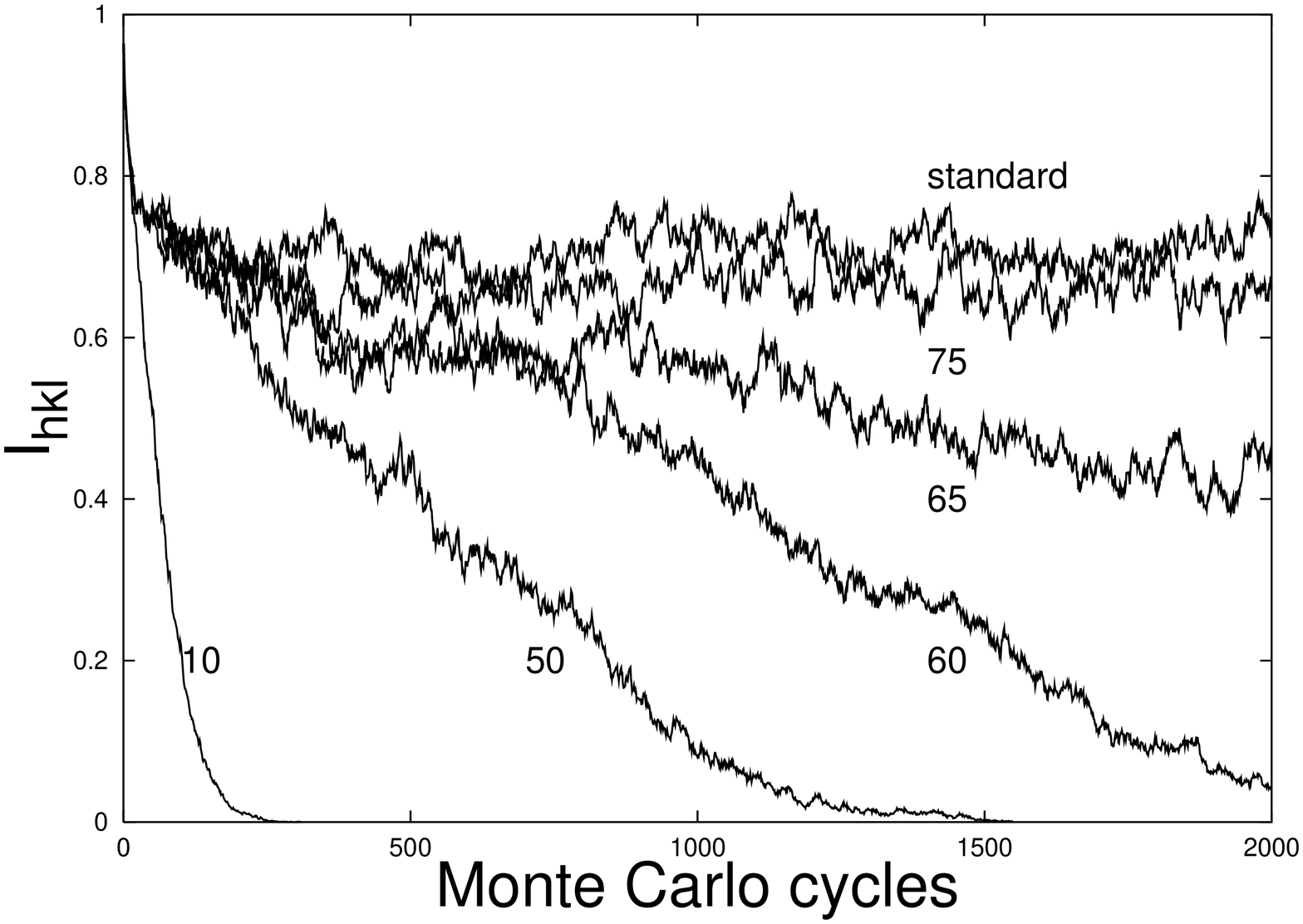}%
 \caption{
}
 \end{figure}
 \begin{figure}[!]
 \includegraphics[height=400pt,width=500pt]{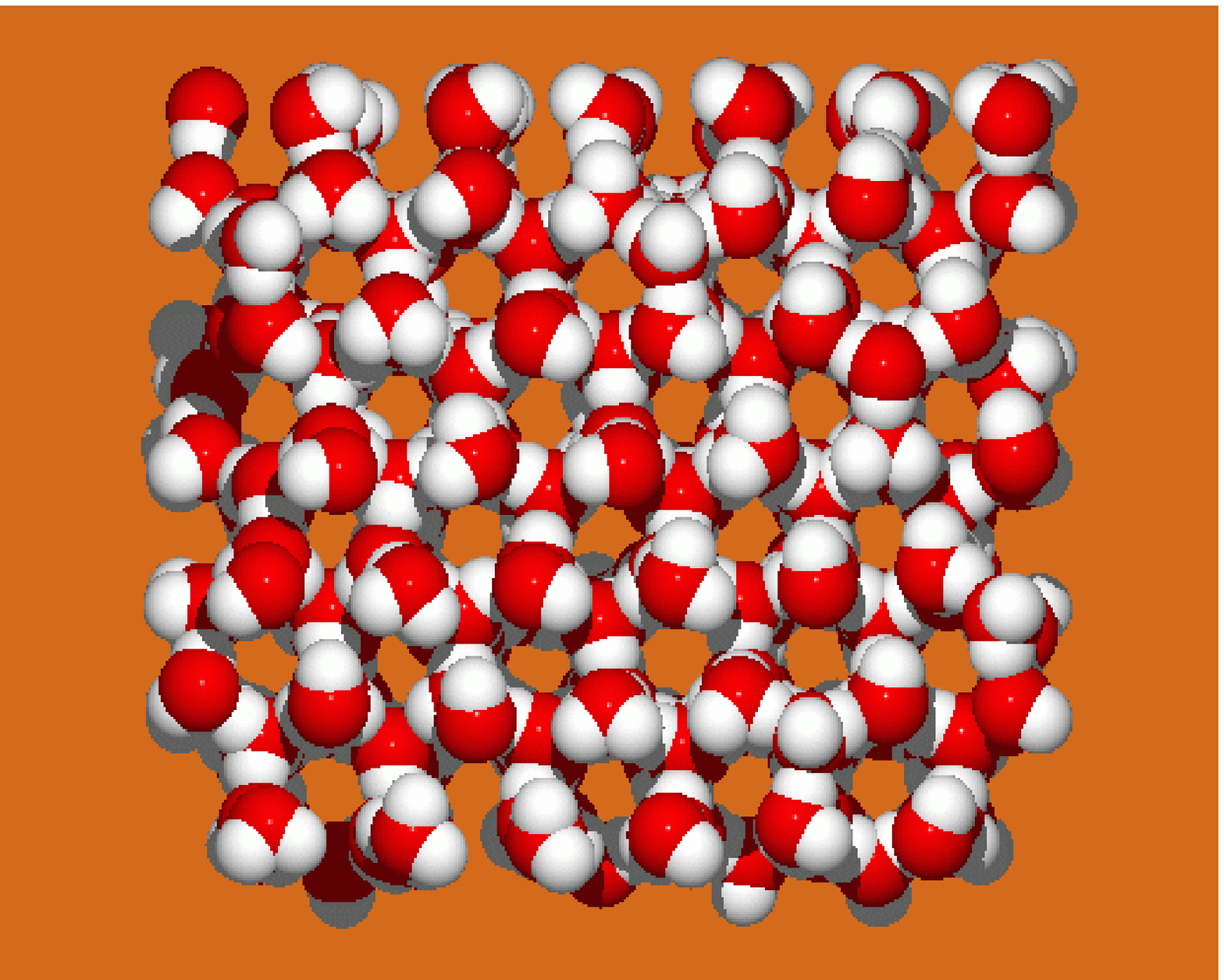}%
 \caption{
}
 \end{figure}
 \begin{figure}[!]
 \includegraphics[height=400pt,width=500pt]{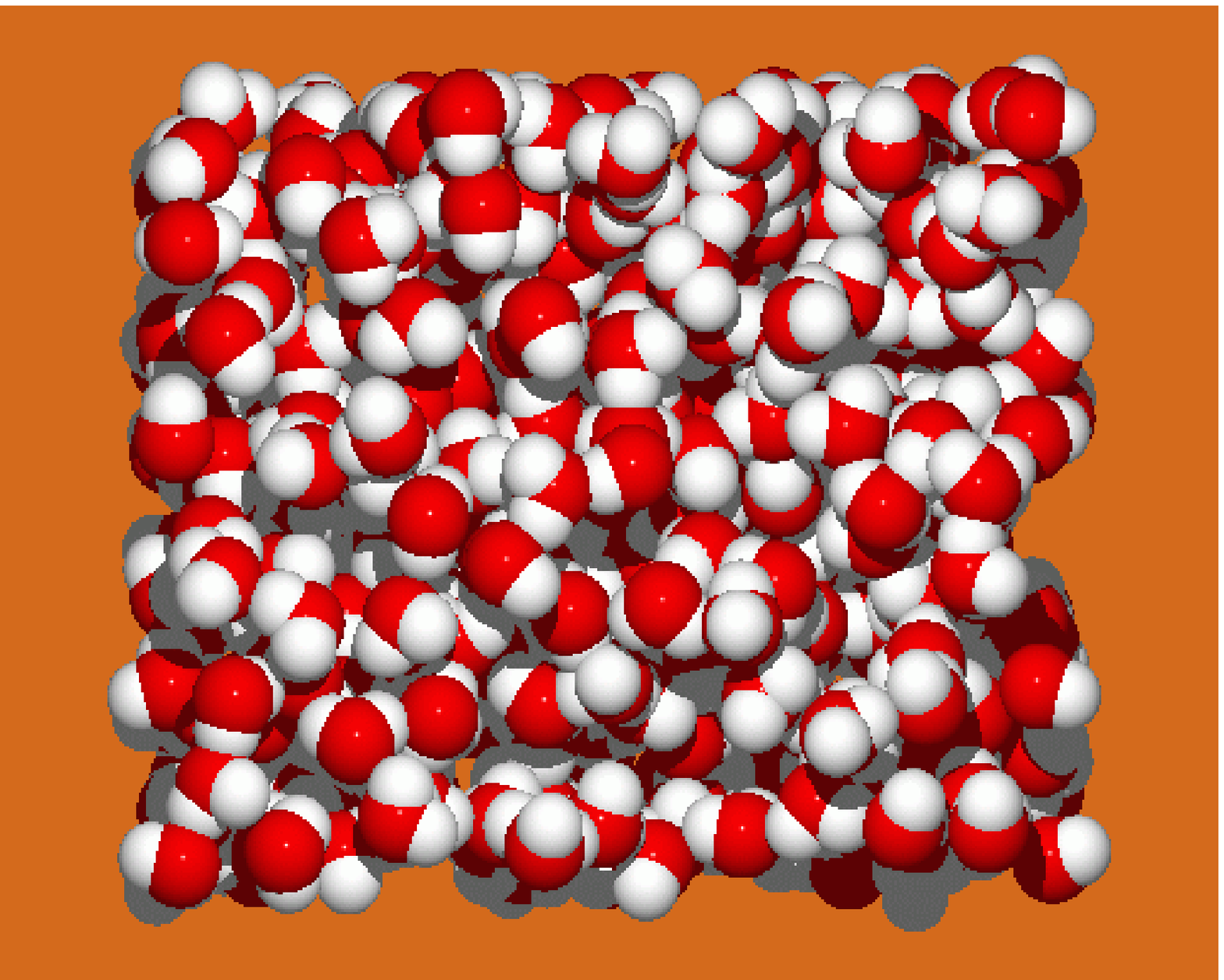}%
 \caption{
}
 \end{figure}
\end{document}